\documentclass[pra,twocolumn,amsmath,amssymb,superscriptaddress]{revtex4}

\usepackage{graphicx}

\def\Per{\mathop{\rm Per}}

\begin{document}

\title{Complexity of full counting statistics of free quantum particles in product states}

\author{Dmitri~A.~Ivanov}
\affiliation{Institute for Theoretical Physics, ETH Z\"urich,
8093 Z\"urich, Switzerland}

\author{Leonid~Gurvits}
\affiliation{Department of Computer Science, The City College of New York,
New York, NY 10031, USA}

\begin{abstract}
We study the computational complexity of quantum-mechanical expectation values of single-particle
operators in bosonic and fermionic multi-particle product states. Such expectation values appear,
in particular, in full-counting-statistics problems. Depending on the initial multi-particle
product state, the expectation values may be either easy to compute (the required number of operations
scales polynomially with the particle number) or hard to compute (at least as hard as a permanent
of a matrix). However, if we only consider full counting statistics in a finite number of
final single-particle states, then the full-counting-statistics generating function becomes
easy to compute in all the analyzed cases. We prove the latter statement for the general case
of the fermionic product state and for the single-boson product state (the same as used in
the boson-sampling proposal). This result may be relevant for using multi-particle product states
as a resource for quantum computing.
\end{abstract}

\date{November 23, 2019}

\maketitle

\section{Introduction}
\label{sec:Introduction}

Future quantum computers are predicted to efficiently solve certain problems
difficult for classical ones \cite{nielsen-2000}. One indication of this
``quantum supremacy'' is the computational complexity of quantum amplitudes:
computationally simple quantum states and operators may generate expectation
values of higher complexity. This consideration lead to a quantum-computing proposal
named ``Boson sampling'' \cite{aaronson-2013}, where bosonic multi-particle amplitudes
are given by (presumably) computationally difficult permanents \cite{valiant-1979}.
In the boson-sampling proposal, the origin of the computational complexity of
the corresponding non-interacting multi-particle amplitudes may be traced down to the quantum
nature of the initial single-boson state. A similar construction with fermions
would require suitably entangled fermionic states, in order to generate scattering
amplitudes of the same complexity level \cite{shchesnovich-2015,ivanov-2017}.

\begin{figure*}
\centerline{\includegraphics[width=.8\textwidth]{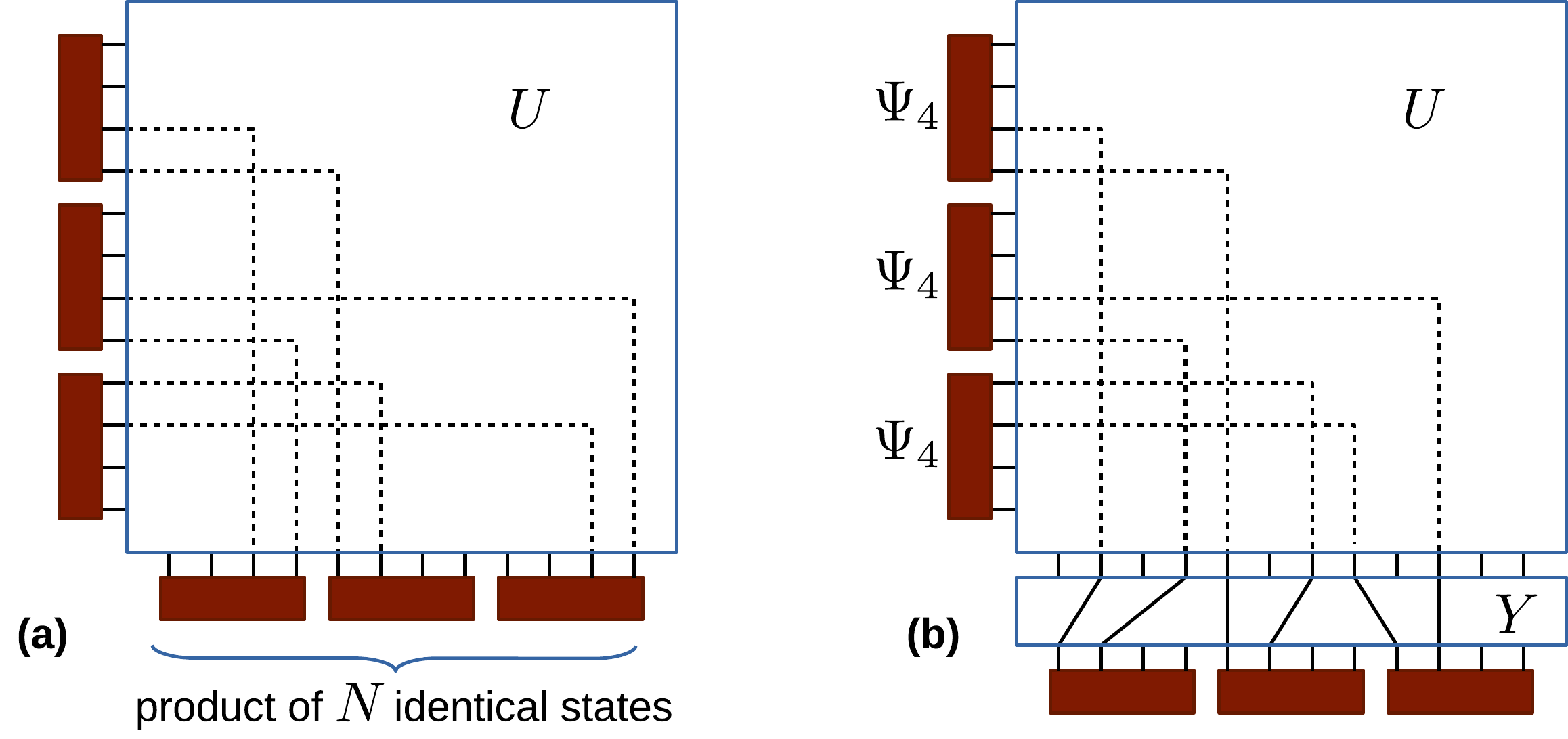}}
\caption{
{\bf (a)} A schematic representation of the matrix element  $\langle \Phi_0 | \hat{U} | \Phi_0 \rangle$
(or  $\langle \Phi_1 | \hat{U} | \Phi_2 \rangle$, if the two states are different). The solid orange rectangles
represent the factors of the product states $| \Phi_0 \rangle$ (or $| \Phi_1 \rangle$ and $| \Phi_2 \rangle$).
The big square represents the single-particle matrix $U$, with the dashed lines representing the matrix elements of $U$
(there are $O(N^2)$ of such matrix elements).
{\bf (b)} The construction used in Section~\ref{subsection-psi4} for proving the equivalence of the hardness of the
quantum amplitudes from Ref.~\onlinecite{ivanov-2017} to the hardness in the definition of the present paper. Here,
all the orange rectangles denote specifically the state $\Psi_4$. The lower rectangle represents the auxiliary
operator $Y$, and the solid lines across this rectangle denote the matrix elements of $Y$ equal to $1$ (with the rest
of the matrix elements being zero).
}
\label{fig1}
\end{figure*}

Those examples suggest that we may benefit from a more systematic study of the complexity
of expectation values for various classes of quantum states and operators.
To some extent, this approach was already developed in the context of quantum
optics \cite{aaronson-2011}, but we find it instructive to discuss the bosonic and fermionic
cases on equal grounds.
Specifically, we restrict our study to the computational
complexity of matrix elements $\langle \Phi_1 | \hat{U} | \Phi_2 \rangle$, where $| \Phi_1 \rangle$ and
$| \Phi_2 \rangle$ are multiparticle bosonic or fermionic states constructed as direct
products of $N\gg 1$ identical states and $\hat{U}$ is a non-interacting
multiparticle operator (e.g., a non-interacting
evolution operator or a similar operator without the unitarity condition;
we use hat for non-interacting multiparticle operators, while the same
letters without hat denote the corresponding single-particle operators, as explained 
in Sec.~\ref{sec:Definitions:operators}), see Fig.~\ref{fig1}a. In this formulation,
the states $| \Phi_i \rangle$ only require a finite number of parameters for their
description (which is automatic in the fermionic case and implies an extra assumption
for bosons) and the operator $\hat{U}$ is defined by the underlying single-particle operator $U$
and is thus parametrized by $O(N^2)$ parameters. We are interested in a criterion
for the matrix element $\langle \Phi_1 | \hat{U} | \Phi_2 \rangle$ to be computable in a polynomial
in $N$ time. In this paper, we only consider the problem of an exact computation
and do not discuss the issue of approximations (the latter may be relevant for
practical quantum-computing applications \cite{aaronson-2011}).

We do not have a full answer to this question, but in this paper we collect a few known examples:
some of them where a polynomial in $N$ algorithm exists and others (specifically, 
the boson-sampling and entangled-fermion examples) that are (at least) as complex as a permanent
(and therefore are believed to belong to a higher complexity class non-computable in
polynomial time).

After this overview of the known results for the general $\hat{U}$,
we consider a variation of the problem where $\hat{U}$
is generated by a single-particle operator $U = 1+V$ with $V$ having a small rank. For the bosonic version
of the problem with the single-product boson state (as in the boson-sampling construction), we find a polynomial
algorithm thus proving Lemma B.5\cite{lemma} of Ref.~\onlinecite{aaronson-2013} presented
there without proof [about the polynomial
computability of the permanent $\Per (1+V)$]. A similar statement for
the fermionic case is also formulated and proven. We also refine the original
formulation of the lemma by proving an estimate for the degree of the polynomial: the
number of the required operations is bounded by $O(N^{2k+1})$ in the bosonic case
and $O(N^{2k})$ in the fermionic case.

The motivation for the above formulations comes partly from the full-counting-statistics
(FCS) problems, where the generating function for the probability distribution of
noninteracting particles has the described structure \cite{levitov-1996}. In particular,
the results for the operators $\widehat{1+V}$ with a finite-rank matrix $V$ correspond
to the computational complexity of the ``marginal'' FCS in a finite number of states
(tracing over the remaining states). We elaborate on this interpretation in the
corresponding section of the paper.

The paper consists of the three main parts. The first part introduces the
multi-particle complexity of product states and reviews, in this context,
previously known results with only minor reformulations. This part includes Section \ref{sec:Definitions}
with definitions and notation and Section \ref{sec:Complexity-general} with examples
for the case of the general noninteracting operator $\hat{U}$. The second part of
the paper is Section \ref{sec:Finite-rank}, where we prove new results for the more
restrictive case $\hat{U}=\widehat{1+V}$. The third part of the paper, Section \ref{sec:FCS},
addresses motivation and interpretation of our constructions in terms
of full counting statistics.
Finally, in Section \ref{sec:Conclusion} we summarize our results
and propose questions for further studies.

\section{Definitions and notation}
\label{sec:Definitions}

\subsection{Fermionic and bosonic states}
\label{sec:Definitions:states}

We consider the fermionic and bosonic multi-particle spaces (Fock spaces)
generated by a large number of single-particle levels, and we are interested
in the computational complexity of matrix elements of a certain class of operators
as a function of this number (whether it is polynomial or higher, e.g., exponential). More specifically,
we restrict our analysis to product states: tensor products of states built on a small number
(one or a few) of single-particle levels. For simplicity, in our discussion we consider all
the states in these products to be identical, even though many of our results may also be extended
to the case of products of different states. The following states will appear in
our examples:
\begin{itemize}
\item {\bf single-boson product state}: 
\begin{equation}
|\mathrm{BN}{=}1\rangle^N = |\mathrm{BN}{=}1\rangle \otimes \ldots \otimes |\mathrm{BN}{=}1\rangle 
\quad
\text{($N$~times),}
\end{equation}
where $|\mathrm{BN}{=}1\rangle = b^\dagger |\star\rangle_{B,1}$ is a single-boson state
($b^\dagger$ here and below denotes the boson creation operator and $|\star\rangle_{B,1}$
is the bosonic vacuum with one empty single-particle level).
This is the state used in the boson-sampling proposal \cite{aaronson-2013}.
\item {\bf coherent-boson product state}:
\begin{equation}
|\mathrm{BC}{=}\alpha\rangle^N = |\mathrm{BC}{=}\alpha\rangle \otimes \ldots \otimes 
|\mathrm{BC}{=}\alpha\rangle 
\quad
\text{($N$~times),}
\end{equation}
where $|\mathrm{BC}{=}\alpha\rangle = \exp(\alpha b^\dagger - \alpha^2/2) |\star\rangle_{B,1}$
is a coherent boson state.
\item {\bf Fermi-sea product state}: a class of states constructed as
\begin{equation}
|\mathrm{FS}\rangle^N=|\mathrm{FS}\rangle \otimes \ldots \otimes |\mathrm{FS}\rangle 
\quad
\text{($N$~times),}
\label{state:FS}
\end{equation}
with $|\mathrm{FS}\rangle = \psi_1^\dagger \ldots \psi_k^\dagger |\star\rangle_{F,n}$, where
$|\star\rangle_{F,n}$ is a fermionic vacuum with $n$ single-particle states and $\psi_i^\dagger$ are
creation operators for some $k\le n$ (mutually orthogonal, for the sake of normalization) linear
combinations of those states. $n$ and $k$ are fixed small numbers (unrelated to $N$). The product
state $|\mathrm{FS}\rangle^N$ then belongs to the multi-particle space (Fock space) generated by $Nn$
single-particle levels. Two particular cases of such a state is the vacuum state ($k=0$) and the fully occupied
state ($k=n$).
\item{\bf entangled-quadruplet product state}:
\begin{equation}
|\Psi_4\rangle^N=|\Psi_4\rangle \otimes \ldots \otimes |\Psi_4\rangle
\quad
\text{($N$~times),}
\label{state:Psi4}
\end{equation}
where $|\Psi_4\rangle=(1/\sqrt{2})(f_1^\dagger f_2^\dagger + f_3^\dagger f_4^\dagger ) |\star\rangle_{F,4}$.
This state was used in Ref.~\onlinecite{ivanov-2017}. It involves $2N$ fermions in $4N$ single-particle states.
\end{itemize}

\subsection{Non-interacting operators}
\label{sec:Definitions:operators}

Every single-particle operator $U$ generates a ``multiplicative'' multi-particle operator $\hat{U}$
in the multi-particle Fock space. A ``physical definition'' of this construction is sometimes written
as
\begin{equation}
\hat{U}=\exp(\sum_{ij} a^\dagger_i (\ln U)_{ij} a_j)\, ,
\label{U-exponent}
\end{equation}
where $a^\dagger_i$ and $a_j$ are either fermionic or bosonic creation and annihilation operators. However,
this definition formally fails when $U$ has zero eigenvalues (non-invertible). For our purpose, we
extend this definition to non-invertible matrices $U$, which can be done either by continuity or with
a more explicit alternative definition
\begin{equation}
\hat{U}a^\dagger_{j_1}\ldots a^\dagger_{j_k}|\star\rangle =
\sum_{i_1\ldots i_k} U_{i_1,j_1} \ldots U_{i_k,j_k} a^\dagger_{i_1}\ldots a^\dagger_{i_k} |\star\rangle\, ,
\label{U-product}
\end{equation}
which describes the action of $\hat{U}$ on each of the basis vectors of the Fock space. 

With this definition, we have a set of the ``non-interacting operators'' $\hat{U}$ defined as those obtainable
from single-particle matrices $U$. This is a representation of the monoid of matrices $U$ with respect
to multiplication (i.e., $\widehat{U_1 U_2}=\hat{U}_1 \hat{U}_2$). In particular, this set is closed with respect
to multiplication. An example of such an operator is a quantum evolution operator for a non-interacting system
of particles (given by (\ref{U-exponent}) with $\ln U$ playing the role of the Hamiltonian).
Another example motivated by full-counting-statistics problems is presented in Section \ref{sec:FCS}
below.

\subsection{Multi-particle complexity of a quantum state (or of a pair of states)}
\label{sec:Definitions:complexity}

Now we are ready to define the main object of our study. We define the {\em multi-particle complexity} of a
pair of states $|\Phi_1\rangle$, $|\Phi_2\rangle$ (or of a single state $|\Phi_0\rangle$)
as the maximal computational complexity of the matrix element
\begin{equation}
\langle \Phi_1 | \hat{U} | \Phi_2 \rangle
\quad
\text{(or~} \langle \Phi_0 | \hat{U} | \Phi_0 \rangle \text{,~respectively)}
\label{matrix-element}
\end{equation}
(with the maximum taken over all non-interacting operators $\hat{U}$),
see Fig.~\ref{fig1}a. The computational complexity is understood
as scaling of the required number of operations as a function of $N$ (see
more explanations in Section~\ref{sec:Definitions:computational} below). The operator
$\hat{U}$ is parametrized by its single-particle counterpart $U$, which requires $N^2$ parameters. The quantum
states $|\Phi_i\rangle$, in their full generality, use an exponential number
of amplitudes, therefore the definition is only meaningful if we restrict it to a subclass of states
parametrized by at most a polynomial in $N$ set of parameters. One possible restriction of this sort is
to consider product states (as defined in Section~\ref{sec:Definitions:states}), where each of the factors
involves only a finite number of parameters (we do not formalize this restriction further).
Note that the operators $U$ generally act across all the factors in the product state, which
may make the matrix element (\ref{matrix-element}) computationally demanding.

In Section~\ref{sec:Finite-rank} below, we also consider a modification of this definition where $\hat{U}$ is
further restricted to be generated by a matrix $U=1+V$, where $V$ is a matrix of a finite rank. We will
call this a {\em finite-rank complexity} of a state (or of a pair of states).

\subsection{Computational complexity for real and complex functions}
\label{sec:Definitions:computational}

Defining computational complexity for functions with continuous variables is sometimes a subtle issue \cite{weihrauch-2000},
and we do not want to go deeply into this topic here. Instead, since the expectation values of interest are
all polynomials of the matrix elements of $U$ and of the wave function components, we define the computational
complexity as the scaling of the number of required arithmetic operations with $N$ (with the exception of the
coherent-boson case, which involves the exponentiation operator, see more details in 
Section \ref{subsec:coherent-boson-complexity}).
To simplify our notation, we only distinguish two levels of complexity: ``easy'' (computable in a polynomial in $N$
number of operations) and ``hard'' (at least as difficult as computing a matrix permanent).

There is a general belief that computing a permanent requires a higher than polynomial number of
operations, which implies $P\ne NP$ \cite{valiant-1979}. We also need this assumption in order for our classification to
be meaningful. However otherwise we never make use of it.

\section{Complexity in case of general $\hat{U}$}
\label{sec:Complexity-general}

We do not have a general criterion for product states to be ``easy'' or ``hard'', but we can give
a few examples of states of each of them:
\begin{itemize}
\item
Single-boson product state is ``hard''.
\item
Coherent-boson product state is ``easy''.
\item
Fermi-sea product state is ``easy''.
\item
Entangled-quadruplet product state is ``hard''.
\end{itemize}

\subsection{Single-boson product state is ``hard''}

The corresponding expectation value is a permanent,
\begin{equation}
\langle\mathrm{BN}{=}1|^N \hat U |\mathrm{BN}{=}1\rangle^N = \Per U\, ,
\end{equation}
so it is ``hard'' by definition. This high complexity was used in Ref.~\onlinecite{aaronson-2013}
to conjecture the ``quantum supremacy'' of Boson Sampling.

\subsection{Coherent-boson product state is ``easy''}
\label{subsec:coherent-boson-complexity}

Since non-interacting operators $\hat{U}$ act within the space of coherent states
(and this action can be written in single-particle terms), one can easily calculate
the matrix element of $\hat{U}$ between any two coherent states. In particular,
\begin{equation}
\langle\mathrm{BC}{=}\alpha|^N \hat{U} |\mathrm{BC}{=}\alpha\rangle^N = 
\exp\left[\alpha^2 \left( \sum_{ij} U_{ij} - N \right) \right]\, .
\end{equation}

In this example, unlike in all the others, we use a sloppy definition
of complexity: instead of the wave-function components (there are infinitely many of them),
we use the parameter $\alpha$ of the coherent state and are allowed one exponentiation
at the end of the calculation.

\subsection{Fermi-sea product state is ``easy''}
\label{subsec:Fermi-product}

The product of Fermi seas (\ref{state:FS}) is also a Fermi sea with $Nk$ fermions.
For this large Fermi sea, one easily finds
\begin{equation}
\langle\mathrm{FS}|^N \hat{U} |\mathrm{FS}\rangle^N = \det_{i,j} \langle\psi_i| U |\psi_j\rangle\, ,
\end{equation}
where the determinant is of the $Nk$-dimensional matrix of the single-particle matrix elements
between the states generating the large Fermi sea. This proves that this matrix element is computable
in polynomial time. Note that this argument equally applies to products of non-identical Fermi seas.

\subsection{Entangled-quadruplet product state is ``hard''}
\label{subsection-psi4}

This was shown in Ref.~\onlinecite{ivanov-2017} 
(it also follows from the results on mixed discriminants in Refs.~\onlinecite{gurvits-2005,gurvits-2009}).
Strictly speaking, in that work, the hardness of $\langle x| \hat{U} |\Psi_4\rangle^N$
was proven, where $|x\rangle$ is a Fermi sea with arbitrary $2N$ states (orthogonal, for simplicity), 
$|x\rangle=\psi_1^\dagger \ldots \psi_{2N}^\dagger |\star\rangle_{F,4N}$. However, we can easily convert this statement
into one for the expectation value in the state $|\Psi_4\rangle^N$. Namely, consider a single-particle operator $Y$
transforming the states $\psi_1$, \ldots, $\psi_{2N}$ into the basis states $f_1$, $f_2$, $f_5$, $f_6$, \ldots,
$f_{4N-3}$, $f_{4N-2}$ (in arbitrary order) and zeroing out the orthogonal complement of $\psi_1$, \ldots, $\psi_{2N}$.
Then (see Fig.~\ref{fig1}b)
\begin{multline}
\langle\Psi_4|^N \hat{Y} \hat{U} |\Psi_4\rangle^N =
\langle\Psi_4|^N \hat{Y} |x\rangle \langle x|\hat{U} |\Psi_4\rangle^N \\
=2^{-N/2} \langle x|\hat{U} |\Psi_4\rangle^N\, ,
\end{multline}
which proves the hardness of the left-hand side of the above equation.

\section{Finite-rank complexity}
\label{sec:Finite-rank}

In this section, we consider the finite-rank complexity:
a modified version of the complexity definition (Section~\ref{sec:Definitions:complexity}),
where the operators $\hat{U}$ are restricted to those generated by
\begin{equation}
U=1+V\, ,
\label{U-finite-rank}
\end{equation}
where $V$ is a matrix of a finite rank:
\begin{equation}
V_{ij}=\sum_{s=1}^k u^{(s)}_i v^{(s)}_j\, .
\label{V-finite-rank}
\end{equation}

Obviously, the finite-rank complexity cannot be higher than the complexity for the general $\hat{U}$.
In particular, for all the examples considered above, the finite-rank complexity is ``easy'' (polynomial).
Moreover, we can prove that the finite-rank complexity is polynomial for a general product state
in the fermionic case. Specifically, we prove the following two statements below:
\begin{itemize}
\item
The finite-rank complexity of the single-boson product state is ``easy''.
We can further prove that the number of required operations scales as $O(N^{2k+1})$.
\item
The finite-rank complexity of {\em any} fermionic product state is ``easy''.
The number of required operations is also limited as $O(N^{2k})$.
\end{itemize}

\subsection{Finite-rank complexity of the single-boson product state is ``easy''}

The matrix element is given by the permanent
\begin{equation}
\langle\mathrm{BN}{=}1|^N \hat U |\mathrm{BN}{=}1\rangle^N = \Per U = \Per (1+V)\, .
\label{finite-rank-permanent}
\end{equation}

Below we show that, if $V$ has a finite rank $k$, the permanent (\ref{finite-rank-permanent})
may be expressed in terms of the coefficients of an auxiliary polynomial of degree $2N$ in $2k$
variables, which, in turn, requires only a polynomial in $N$ number of operations.

A simple combinatorial argument expresses the permanent (\ref{finite-rank-permanent})
in terms of the vectors $u^{(s)}$ and $v^{(s)}$
from Eq.~(\ref{V-finite-rank}):
\begin{multline}
\Per (1+V) \\
=  \sum_{X\subseteq \{1,\ldots,N\} }
\sum_{\substack{
   s_u(X),s_v(X) \\
   [s_u(X)]=[s_v(X)]
}} \prod_{x\in X}
u^{(s_u(x))}_x v^{(s_v(x))}_x
\prod_{r=1}^k n_r! \, ,
\label{per-finite-rank-expansion-1}
\end{multline}
where the first sum is taken over all subsets $X$ of the set of indices $\{1,\ldots,N\}$, the second sum is over
the label sets $s_u$ and $s_v$ (ranging from $1$ to the rank $k$) for elements of $X$ such that they form
identical multisets (sets with repetitions) $[s_u(X)]=[s_v(X)]$, but possibly permuted with respect to each other.
Finally, in the last product $n_r$ denotes the multiplicity of $r$ in the multiset $[s_u(X)]$
(or, equivalently $[s_v(X)]$). This expression may, in turn, be computed with the help of the auxiliary polynomial
of $2k$ formal variables
\begin{multline}
F(a_u^{(1)},\ldots a_u^{(k)}, a_v^{(1)},\ldots,a_v^{(k)}) \\
= \prod_{x=1}^N
\left[1+
\sum_{s=1}^k \sum_{s'=1}^k
a_u^{(s)}  a_v^{(s')}  u^{(s)}_x v^{(s')}_x 
\right] \\
= \sum_{\{n_r\},\{n'_r\}} F_{n_1,\ldots,n_k,n'_1,\ldots,n'_k} \\
\times\; (a_u^{(1)})^{n_1} \ldots (a_u^{(k)})^{n_k} (a_v^{(1)})^{n'_1} \ldots (a_v^{(k)})^{n'_k} \, ,
\label{per-finite-rank-polynomial}
\end{multline}
where the first equality is the definition of the polynomial $F(a_u^{(1)},\ldots a_u^{(k)}, a_v^{(1)},\ldots,a_v^{(k)})$
and the second equality is its expansion in powers of $a_u^{(s)}$ and $a_v^{(s)}$ defining its coefficients.
On inspection, the ``diagonal'' coefficients of this polynomial reproduce the terms in the sum 
(\ref{per-finite-rank-expansion-1}), up to combinatorial coefficients, and one finds
\begin{equation}
\Per (1+V) = \sum_{\{n_r\}}
F_{n_1,\ldots,n_k,n_1,\ldots,n_k} \prod_{r=1}^k n_r!\, .
\label{per-finite-rank-solution}
\end{equation}
There are altogether $O(N^{2k})$ coefficients $F_{n_1,\ldots,n_k,n'_1,\ldots,n'_k}$, including
$O(N^k)$ diagonal coefficients (with $n_r=n'_r$). Their calculation involves multiplying out
$N$ terms in Eq.~(\ref{per-finite-rank-polynomial}), where at each multiplication the $O(N^{2k})$
coefficients need to be updated. Therefore the calculation of $\Per (1+V)$ using
Eqs.\ (\ref{per-finite-rank-polynomial}) and (\ref{per-finite-rank-solution})
can be done in $O(N^{2k+1})$ operations, as claimed. This proves
Lemma B.5\cite{lemma} of Ref.~\onlinecite{aaronson-2013}
and the ``finite-rank easiness'' of the single-boson product state.

\subsection{Finite-rank complexity of any fermionic product state is ``easy''}

The idea of the proof is that $\hat{U}$, in the finite-rank construction
(\ref{U-finite-rank})--(\ref{V-finite-rank}), acts nontrivially only in a small
subspace spanned by a small number of fermionic states and therefore may be written in
terms of a small number of fermionic operators. Specifically, $\hat{U}$ may be written
in terms of the creation and annihilation operators defined as
\begin{equation}
\hat{u}^\dagger_s=\sum_i u^{(s)}_i f^\dagger_i\, ,
\qquad
\hat{v}_s= \sum_j v^{(s)}_j f_j\, ,
\end{equation}
where $f^\dagger_i$ and $f_j$ are the fermionic creation and annihilation operators
in the original basis.
Using the definition (\ref{U-product}), one can verify that $\hat{U}$ may be expressed
as the polynomial in those operators,
\begin{equation}
\hat{U}=\sum_{ \{s_i\} }
\hat{u}^\dagger_{s_1} \ldots \hat{u}^\dagger_{s_r}
\hat{v}_{s_r} \ldots \hat{v}_{s_1}\, ,
\label{U-polynomial}
\end{equation}
where the sum is taken over all subsets $\{s_i\}$ of indices (including the empty subset, which contributes
the unity operator) and $r$ is the number of elements in the subset.
The polynomial (\ref{U-polynomial}) has $2^k$ terms with its degree limited by $r\le k$.

Now consider the expectation value of each term of the polynomial (\ref{U-polynomial}) in any
product state
\begin{equation}
|\Phi_0\rangle=|\Psi_{(1)}\rangle \otimes \ldots |\Psi_{(N)}\rangle\, ,
\label{fermi-product}
\end{equation}
where each of the states $|\Psi_{(i)}\rangle$ belongs to a Fock space generated by a ``small''
(not growing with $N$) number 
of single-particle states (for our argument, we do not even need these states to be identical).
Our states (\ref{state:FS}) and (\ref{state:Psi4}) are particular cases of this construction.
Without loss of generality, we may take the states $|\Psi_{(i)}\rangle$ to be normalized.

To calculate the expectation value of a term of degree $r$ in the polynomial (\ref{U-polynomial}) in
the state $|\Phi_0\rangle$,
we decompose each of the operators $\hat{u}^\dagger_s$ and $\hat{v}_s$ into $N$ components:
\begin{equation}
\hat{u}^\dagger_s = (\hat{u}^\dagger_s)_1 \oplus \ldots \oplus (\hat{u}^\dagger_s)_N\, ,
\end{equation}
where $(\hat{u}^\dagger_s)_i$ acts in the $i$-th space (hosting the state $|\Psi_{(i)}\rangle$).
The same decomposition is done for  the operators $\hat{v}_s$. Now we expand the product of $2r$ operators
in Eq.~(\ref{U-polynomial}) to obtain a sum $N^{2r}$ terms. Each of these terms
is itself a product of $N$ factors by the number of subspaces in Eq.~(\ref{fermi-product}).
These factors have the form 
$\langle \Psi_{(i)}|(\hat{u}^\dagger_{j_1})_i \ldots (\hat{u}^\dagger_{j_m})_i 
(\hat{v}_{j'_1})_i \ldots (\hat{v}_{{j'}_{m'}})_i |\Psi_{(i)}\rangle$: they represent an expectation value
in a ``small'' (not growing with $N$) space and therefore can be computable in a ``small'' number of
operations. Moreover, at most $2k$ of those expectation values are nontrivial, and the rest are
equal to one, since we have taken $|\Psi_{(i)}\rangle$ to be normalized. Therefore, the
product of the $N$ factors can actually be computed in a ``small'' (not growing in $N$) number of
operations. Since there are $N^{2r}$ such products for each term of degree $r\le k$ in the polynomial
(\ref{U-polynomial}), we can compute the expectation value $\langle \Phi_0|\hat{U} |\Phi_0\rangle$
in  $O(N^{2k})$ operations. This proves our statement.

\section{Implications for full counting statistics}
\label{sec:FCS}

\begin{figure*}
\centerline{\includegraphics[width=.65\textwidth]{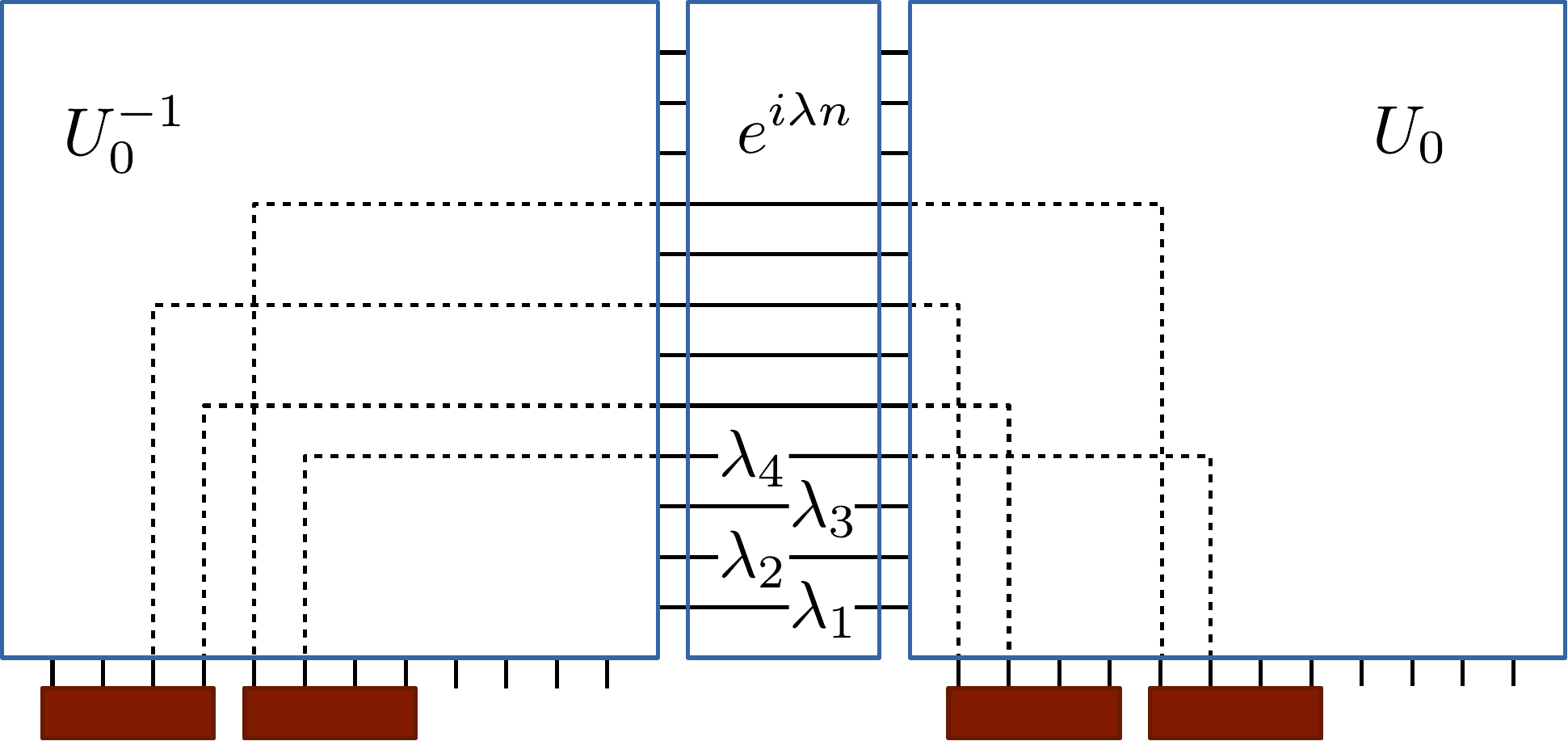}}
\caption{
A schematic illustration of the full-counting-statistics generating function
$\chi(\lambda_1,\ldots,\lambda_N)=\langle \Phi_0 | \hat{U} | \Phi_0 \rangle$,
where $U$ is given by Eq.~(\ref{fcs-eqn}). 
The solid orange rectangles represent the factors of the product states $| \Phi_0 \rangle$.
The big squares are the factors $U_0^{-1}$ and $U_0$ in Eq.~(\ref{fcs-eqn}).
The rectangle in the middle is the factor $e^{i\sum \lambda_i n_i}$. The horizontal lines
marked with $\lambda_i$ correspond to general values of the parameters $\lambda_i$, solid
lines to $\lambda_i=0$, missing lines to $\lambda_i \to i\infty$.
}
\label{fig2}
\end{figure*}

\subsection{Generating function for the particle-number probability distribution}

The above discussion of the complexity of the expectation values may be interpreted in the
language of so called full counting statistics (FCS): a class of problems addressing the probability
distribution of a quantum observable \cite{levitov-1996}. Namely, our results may be reformulated in terms
of complexity of FCS generating functions for non-interacting particles initially prepared
in a certain state $| \Phi_0 \rangle$. Indeed, consider an initial state $| \Phi_0 \rangle$
that is subject to a non-interacting evolution $\hat{U}_0$. We may further define the generating function
\begin{equation}
\chi(\lambda_1,\ldots,\lambda_N)=\sum_{\{n_i\}} e^{i \sum_{i=1}^N \lambda_i n_i}
P(n_1,\ldots,n_N)\, ,
\label{chi-1}
\end{equation}
where 
\begin{equation}
P(n_1,\ldots,n_N)= \left| \langle n_1,\ldots, n_N | \hat{U}_0 | \Phi_0 \rangle \right|^2
\label{chi-2}
\end{equation}
is the probability to observe the counts $n_i$ in the single-particle states $i$ after the evolution $\hat{U}_0$.

In fact, there are three commonly used formulations for the full-counting-statistics problem: one may be
interested either in computing the probabilities (\ref{chi-2}) or in the generating function (\ref{chi-1})
or in sampling the probability distribution with a randomized algorithm. The translation between these
three formulations may turn out to be computationally intensive in case of large $N$. For the purpose
of this paper, we only consider the problem of calculating the generating function (\ref{chi-1})
and do not discuss its connections to the other two formulations (except for a short remark at the
end of Section \ref{sec:fcs-finite-rank}).

The generating function (\ref{chi-1}) has the required structure 
$\chi(\lambda_1,\ldots,\lambda_N)=\langle \Phi_0 | \hat{U} | \Phi_0 \rangle$, where
\begin{equation}
U=U_0^{-1} e^{i \sum_{i=1}^N \lambda_i n_i} U_0
\label{fcs-eqn}
\end{equation}
and $n_i$ is the single-particle projector on the state $i$ (see Fig.~\ref{fig2}).

Some of the parameters $\lambda_i$ used for counting particles in different single-particle states may
be set to zero: in this case, the corresponding particle numbers are simply ignored (a trace is taken over
all the particle numbers). Alternatively, it is also possible to directly take the limit $\lambda_i \to i\infty$
(or, equivalently, $e^{i\lambda_i} \to 0$, which corresponds to projecting onto the states
with zero occupancy of the corresponding single-particle state (in this case, the operator (\ref{fcs-eqn})
is no longer unitary).

At the same time, the state  $| \Phi_0 \rangle$ may be allowed to span only a subset of the available single-particle
input states. Together with the possibility to exclude some output states with the $\lambda_i \to i\infty$ limit,
this provides a lot of flexibility for constructing the operator $U$, without the constraint of unitarity.
Therefore, we conjecture that our results from Section 
\ref{sec:Complexity-general} literally translate into the computational complexity of the generating function
(\ref{chi-1}), (\ref{chi-2}) for the general choice of the non-interacting evolution $U_0$ and
of the complex variables $\lambda_i$.

\subsection{Full counting statistics in a small subset of states}
\label{sec:fcs-finite-rank}

The discussion above applies to the case of the general choice of the parameters $\lambda_i$ for a large
(of order $N$) number of states. We may however consider a simpler problem with only a small number of
nonzero $\lambda_i$ (while keeping the total number of single-particle states and particles large, of order $N$).
Then the matrix (\ref{fcs-eqn}) has the form $1+V$, where $V$ has a finite rank. Indeed, we can rewrite (\ref{fcs-eqn}) as
\begin{equation}
U=1+U_0^{-1} (1-e^{i \sum_{i=1}^N \lambda_i n_i}) U_0\, ,
\end{equation}
and the second term has a finite rank for a finite number of nonzero $\lambda_i$.

Therefore, our results of Section \ref{sec:Finite-rank} fully translate into the statement about
full counting statistics. Namely, in our setup, for a finite subset of states, the
FCS generating function is computable in polynomial time for the states $ | \Phi_0 \rangle$ considered
in Section \ref{sec:Finite-rank} (single-boson product state and any fermionic product states).

Note that for a small subset of states, there is no difference between the three different
formulations of the FCS, since the generating function (\ref{chi-1}) and probabilities (\ref{chi-2})
are related by a Fourier transform computable in a polynomial (in $N$) time in this case. Furthermore,
the sampling can also be performed in a polynomial time in our examples, since there is only a
polynomial in $N$ number of probabilities (\ref{chi-2}).

\section{Summary and discussion}
\label{sec:Conclusion}

The purpose of this paper is two-fold. First, we introduce the notion of the ``multi-particle complexity''
of product states. This definition naturally leads to the question of formulating a criterion for a product
state to be ``hard''.
From examples, one may conjecture that most of such states are actually ``hard'' except for a few special cases. 
One such special case are so called Gaussian states, where Wick theorem applies (see, e.g.,
Ref.~\onlinecite{wang-2007} for definition in the bosonic case). Our examples of coherent-boson and
Fermi-sea product states belong to this class of Gaussian states. One finds more Gaussian states among the mixed
states described by a density matrix, but in this paper we restrict our discussion to pure states only. We
do not know if there is any non-Gaussian state that would generate ``easy'' product states.

Another question in connection with this ``multi-particle complexity'' concept is its possible implications
for quantum computing. Ref.~\onlinecite{aaronson-2013} suggests that this setup (specifically, the example
of Boson sampling) is insufficient for universal quantum computing, but, to our knowledge, without
solid justification. In any case, it would be an interesting problem to characterize the class of problems solvable
in polynomial time with this ``full counting statistics'' setup: an initial preparation of a certain product
state (e.g., one of our ``hard'' examples), then evolution with a single-particle operator (which encodes the
``quantum algorithm''), and finally a measurement of a certain generating function (or of a set of
generating functions) (\ref{chi-1}). It seems plausible that all ``hard'' quantum states are equivalent
for this quantum-computing setup, and therefore all of them would be equivalent to Boson sampling.

In this context, we would like to comment on the relation of our ``multi-particle complexity'' examples
to earlier studies on quantum computing with free bosons and fermions.
In Ref.~\onlinecite{terhal-2002}, computations with non-interacting operators on fermionic systems
were shown to be ``easy''. This is consistent with our examples and with our conjecture above, since
Ref.~\onlinecite{terhal-2002} only considered initial states in the form of ``bitstrings'', which
falls in the category of Gaussian states (our Fermi-sea example in Section~\ref{subsec:Fermi-product}).
It is the choice of the initial state that allows computationally ``hard'' expectation values in our examples.
Note that Ref.~\onlinecite{terhal-2002} considered a more general form of quadratic operators possibly
including pair creation and annihilation terms, while we only restrict our discussion to operators
conserving the particle number.  In Ref.~\onlinecite{knill-2001}, the authors have shown that free
bosons may be used for efficient quantum computing. This is again consistent with our examples, since
Ref.~\onlinecite{knill-2001} uses single-boson states, which 
provides the key ingredient for creating quantum amplitudes of high computational complexity.

We would like to remind the reader that the ``hardness'' of a matrix element
$\langle \Phi_0 | \hat{U} | \Phi_0 \rangle$ does not imply a possibility to actually compute this
quantity with a quantum system. There are two reasons for this. First, the quantum measurement
implies sampling, and achieving a good precision in a typically exponentially small expectation value
would require exponentially many repeated measurements. Second, in this paper we only address
the question of an exact computation, while for experimental implications it may be more relevant to
study approximations. Computational complexity of
approximate computations of permanents, mixed discriminants and other related functions is
addressed in many recent works \cite{gurvits-2005,gurvits-2009,aaronson-2011,aaronson-2014,anari-2017}.

The second goal of the paper is to report two results related to the ``finite-rank'' full counting statistics.
For the cases we managed to prove (any fermionic product states and the single-boson product state), we have
shown that counting particles in a {\em finite} number of final states is an ``easy'' task (computable in
polynomial time).
It seems plausible that this statement might be extended to a wider class of bosonic states (e.g., to any
bosonic product states based on states with a finite number of particles). We leave this extension for future
studies.




\begin{thebibliography}{99}

\bibitem{nielsen-2000}
M.~A.~Nielsen and I.~L.~Chuang,
{\it Quantum Computation and Quantum Information}
(Cambridge University Press, 2000).

\bibitem{aaronson-2013}
S.~Aaronson and A.~Arkhipov,
{\it The computational complexity of linear optics},
Theory of Computing, {\bf 9}, 143 (2013).

\bibitem{valiant-1979}
L.~G.~Valiant,
{\it  The complexity of computing the permanent},
Theor.\ Comp.\ Sci.\ {\bf 8}, 189 (1979).

\bibitem{shchesnovich-2015}
V.~S.~Shchesnovich,
{\it Boson-Sampling with non-interacting fermions},
Int.\ J.\ Quantum Inform., {\bf 13}, 1550013 (2015).

\bibitem{ivanov-2017}
D.~A.~Ivanov,
{\it Computational complexity of exterior products
and multiparticle amplitudes of noninteracting fermions
in entangled states},
Phys.\ Rev.\ A {\bf 96}, 012322 (2017).

\bibitem{aaronson-2011}
S.~Aaronson,
{\it A linear-optical proof that the permanent is \#P-hard},
e-print arXiv:1109.1674 (2011).

\bibitem{lemma}
The same lemma has number 67 in the arXiv version of Ref.~\onlinecite{aaronson-2013}
and number 68 in the manuscript version on the S.~Aaronson's website.

\bibitem{levitov-1996}
 L.~S.~Levitov, H.-W.~Lee, and G.~B.~Lesovik,
{\it Electron Counting Statistics and Coherent States of Electric Current},
J.\ Math.\ Phys.\ {\bf 37}, 4845 (1996).

\bibitem{weihrauch-2000}
K.~Weihrauch,
{\it Computable analysis, an introduction}
(Springer, 2000).

\bibitem{gurvits-2005}
L.~Gurvits,
{\it On the complexity of mixed discriminants and related problems},
in {\it Mathematical Foundations of Computer Science}, {\bf 3618}, 447 (Springer, 2005).

\bibitem{gurvits-2009}
L.~Gurvits,
{\it On complexity of the mixed volume of parallelograms},
in {\it 25th European Workshop on Computational Geometry}, 
{\tt http://2009.eurocg.org}, 337, Brussels, Belgium (2009).

\bibitem{wang-2007}
X.-B.~Wang, T.~Hiroshima, A.~Tomita, and M.~Hayashi,
{\it Quantum information with Gaussian states},
Phys.~Rep.\ {\bf 448}, 1 (2007).

\bibitem{terhal-2002}
B.~M.~Terhal and D.~P.~DiVincenzo,
{\it Classical simulation of noninteracting-fermion quantum circuits},
Phys.\ Rev.\ A {\bf 65}, 032325 (2002).

\bibitem{knill-2001}
E.~Knill, R.~Laflamme, and G.~J.~Milburn,
{\it A scheme for efficient quantum computation with linear optics},
Nature {\bf 409}, 46 (2001).

\bibitem{aaronson-2014}
S.~Aaronson and T.~Hance,
{\it Generalizing and Derandomizing Gurvits' Approximation Algorithm for the Permanent}, 
Quantum Inform.\ and Comp., {\bf 14}, 541 (2014).

\bibitem{anari-2017}
N.~Anari, L.~Gurvits, S.~Oveis Gharan, and A.~Saberi,
{\it Simply exponential approximation of the permanent of positive semidefinite matrices},
in {\it the 58th Annual Symposium on Foundations of Computer Science (FOCS 2017)}, 914 (2017).



\end{thebibliography}
\end{document}